# Performance Comparison of some Synchronous Adders

P. Balasubramanian

School of Computer Science and Engineering

Nanyang Technological University

Singapore 639798

## Abstract

This technical note compares the performance of some synchronous adders which correspond to the following architectures: i) ripple carry adder (RCA), ii) recursive carry lookahead adder (RCLA), iii) hybrid RCLA-RCA with the RCA used in the least significant adder bit positions, iv) block carry lookahead adder (BCLA), v) hybrid BCLA-RCA with the RCA used in the least significant adder bit positions, and vi) non-uniform input partitioned carry select adders (CSLAs) without and with the binary to excess-1 code (BEC) converter. The 32-bit addition was considered as an example operation. The adder architectures mentioned were implemented by targeting a typical case PVT specification (high threshold voltage, supply voltage of 1.05V and operating temperature of 25 degrees Celsius) of the Synopsys 32/28nm CMOS technology. The comparison leads to the following observations: i) the hybrid CCLA-RCA is preferable to the other adders in terms of the speed, the power-delay product, and the energy-delay product, ii) the non-uniform input partitioned CSLA without the BEC converter is preferable to the other adders in terms of the area-delay product, and iii) the RCA incorporating the full adder present in the standard digital cell library is preferable to the other adders in terms of the power-delay-area product.

## Keywords

Digital circuits; adder; ripple carry adder; carry lookahead adder; carry select adder; ASIC; standard cells; CMOS.

## 1. Introduction

Adder is an important datapath component of any general purpose micro-processing or digital signal processing unit. Adder is predominantly used in many operations such as multiplication, division, address computation for cache and memory accesses etc. Hence, the performance of an adder carries significant potential for high performance applications involving digital circuits or systems. Many adder architectures have been proposed and many adder designs have also been presented in the literature. In this brief technical correspondence, we consider four homogeneous adder architectures viz. the RCA, the RCLA, the BCLA and the CSLA, and two hybrid adder architectures viz. the RCLA-RCA and the BCLA-RCA to make a performance comparison. This comparison would be useful since the performance of the individual adder architectures were alone analyzed in some works such as [1], [2] with no comparison to the other adder architectures. As a result, the observations made in [1] or [2] may be myopic in nature. Hence, in this technical correspondence, we consider a variety of adder architectures to arrive at some general conclusions, which may serve as a useful reference. We shall not describe the various adder architectures in this technical note but shall cite some relevant references where the corresponding block level and gate level schematics are presented and described. We shall primarily restrict our attention to a brief highlight of the homogeneous and heterogeneous adder architectures considered and then present the individual and joint design metrics estimated and derive the inferences. We also suggest some scope for further work.

## 2. RCA Architectures

RCA architectures can be constructed by cascading single-bit full adders [3 – 7] or dual-bit full adders [8]. For example, a 32-bit RCA can be constructed by cascading 32 full adders (FAs) or 16 dual-bit full adders (DBFAs). The (single-bit) FA adds an augend bit to an addend bit considering any carry input bit and produces the sum bit and any carry overflow bit. The carry output logic of a full adder is equivalent to the majority voter logic used in a triple-modular redundant circuit or system [9 – 11]. On the other hand, the DBFA adds a pair of augend bits to a pair of addend bits considering any carry input bit and produces a pair of sum bits and any carry overflow bit. The RCA architectures comprising FAs and DBFAs are shown in Figures 1a and 1b respectively, where the carry input may be preset to 0.

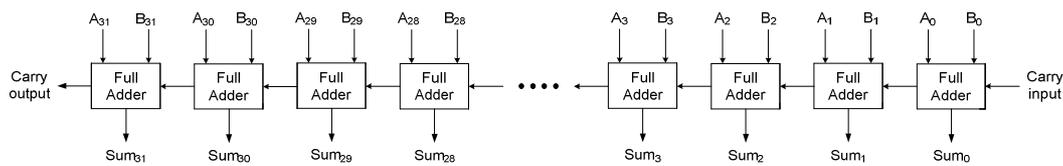

Figure 1a. 32-bit RCA comprising (single-bit) full adders

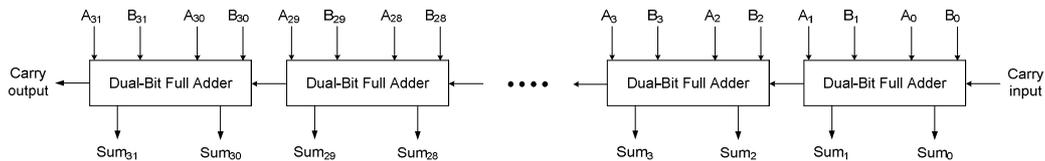

Figure 1b. 32-bit RCA comprising dual-bit full adders (DBFAs)

## 3. CSLA Architectures

The CSLA architecture is constructed by partitioning the input bits and segmenting the addition into many sub-additions [12]. Two types of CSLA architectures are common: i) using a RCA in the least significant adder bit positions and using dual RCAs of appropriate size as dictated by the input partitions, one with a fixed carry input of 0 and another with a fixed carry input of 1. The outputs of the dual RCAs are given to 2:1 multiplexers (MUXes) with the carry output of the preceding input partition serving as the select input for the MUXes of the current input partition, and ii) using a RCA for the least significant adder bit positions, and using a RCA of appropriate size as dictated by the input partitions with a fixed carry input of 0 and the outputs of these RCAs are given to a BEC converter [13] which increments the outputs of the RCAs by 1. The selection of either the outputs of the RCAs with a fixed carry input of 0 or the outputs of the BEC converters is done through the MUXes, which have the carry output from a preceding input partition serving as the select input [1, 14]. The block schematic of the CSLA architecture is portrayed by Figure 2a. Example 4-bit sub-CSLAs with one involving full adders and 2:1 MUXes, and the other involving full adders, 2:1 MUXes and the BEC converters are shown in Figures 2b and 2c respectively. The internal gate-level detail of an example 5-bit BEC is also shown in Figure 2c.

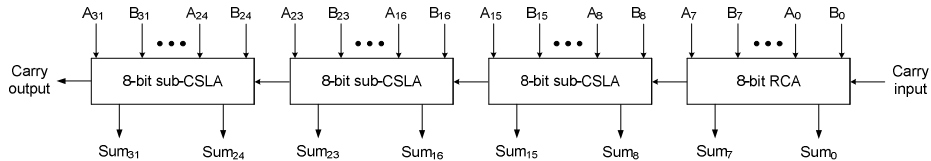

**Figure 2a. 32-bit CSLA architecture**

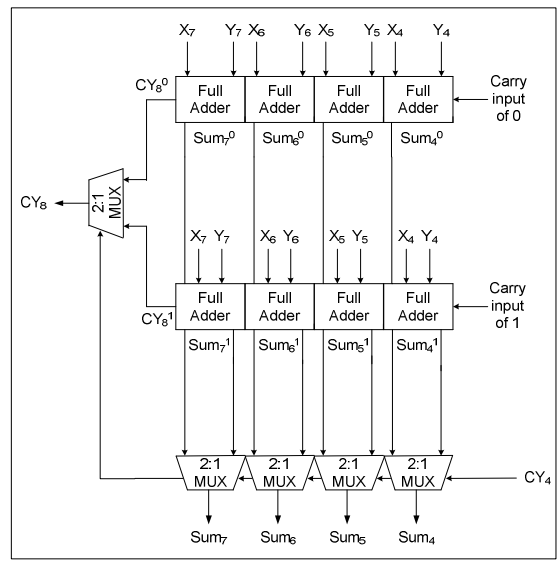

**Figure 2b. Example 4-bit sub-CSLA comprising full adders and 2:1 multiplexers (MUXes)**

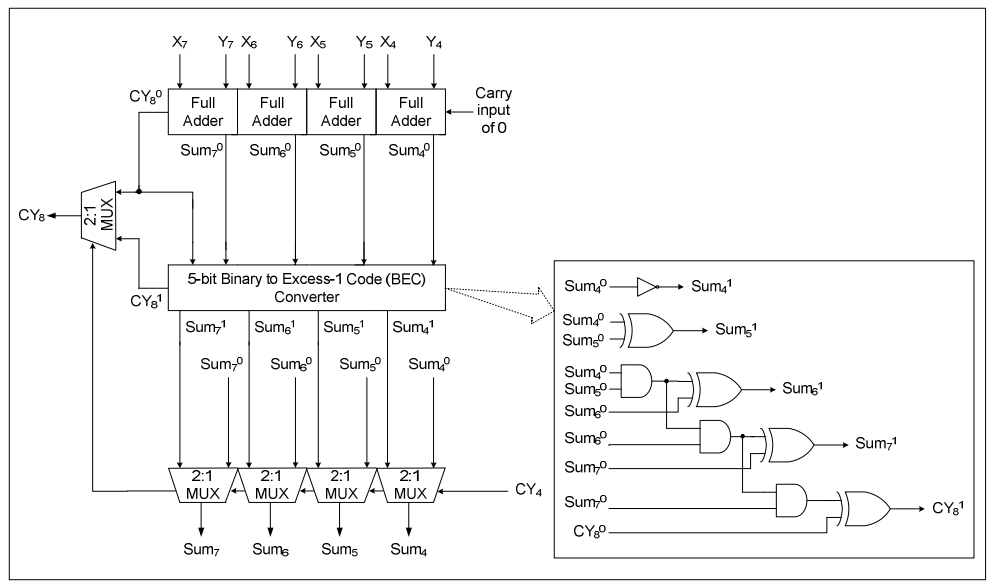

**Figure 2c. Example 4-bit sub-CSLA comprising full adders, 2:1 MUXes and 5-bit BEC converter**

# 4. CLA and CLA-RCA Architectures

## 4.1. RCLA and RCLA-RCA Architectures

The CLA architecture is mainly based on a physical realization of the recursive carry lookahead equations [15]. Different realizations of the RCLA are possible, as given in [16 – 18].

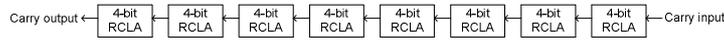

**Figure 3a. Homogeneous RCLA**

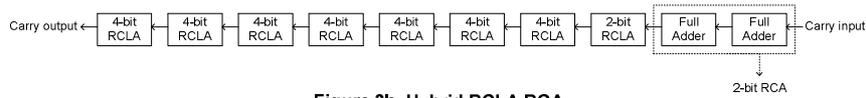

**Figure 3b. Hybrid RCLA-RCA**

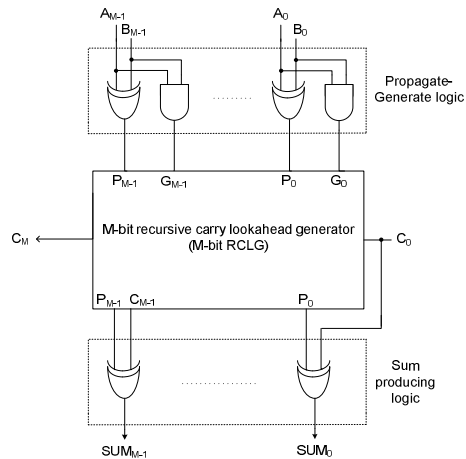

**Figure 3c. Block schematic of M-bit RCLA**

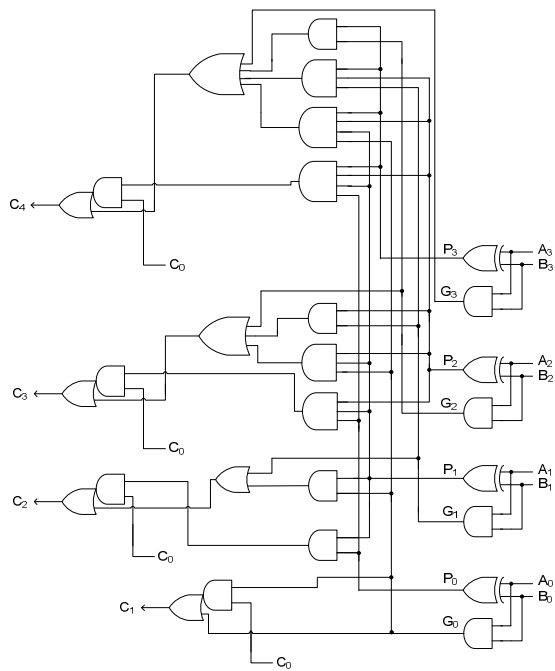

**Figure 3d. Gate-level detail of delay-optimized 4-bit RCLG**

A homogeneous RCLA is constructed using only sub-RCLAs, and a hybrid RCLA-RCA is constructed by using a RCA in the least significant adder bit positions and using sub-RCLAs in the more significant adder bit positions. The block schematics of a delay-optimized 32-bit homogeneous RCLA and a delay-optimized 32-bit hybrid RCLA-RCA are shown in Figures 3a and 3b respectively. The block schematic of a M-bit RCLA (also called the sub-RCLA) is shown in Figure 3c, which consists of the propagate-generate logic, a M-bit recursive carry lookahead generator (RCLG), and the sum producing logic. The gate-level detail of an example delay-optimized 4-bit RCLG is shown in Figure 3d. When this 4-bit RCLG is present in an intermediate adder stage, its critical path will involve just one gate viz. the AO21 complex gate.

## 4.2. BCLA and BCLA-RCA Architectures

The BCLA [19], also called the section-carry based carry lookahead adder (SCBCLA) [17, 18], is another form of the generic CLA, which is also based on the recursive carry lookahead equation and is constructed using sub-BCLAs. However, the BCLA is different from the RCLA. An M-bit BCLA (also called the sub-BCLA) receives a lookahead carry input from a preceding sub-BCLA, as shown in Figure 4c, and produces the lookahead carry output for the successive sub-BCLA. The M-bit BCLA consists of the propagate-generate logic, a M-bit BCLG, and the sum producing logic. The carry input received by the M-bit BCLA and the corresponding augend and addend input bits are serially processed by a cascade of (M–3) full adders and one 3-input XOR gate (resembling a sub-RCA) to produce the corresponding sum output bits. The gate-level detail of a 4-bit BCLG is shown in Figure 4d. Contrary to a M-bit RCLA which produces M lookahead carry outputs, an M-bit BCLA produces only one lookahead carry output. Figure 4a shows the homogeneous BCLA architecture and Figure 4b shows the hybrid BCLA-RCA architecture. A delay-optimized 32-bit BCLA is shown in Figure 4a and a delay-optimized 32-bit hybrid BCLA-RCA is shown in Figure 4b. When the example 4-bit BCLG depicted by Figure 4d is used in an intermediate adder stage, the critical path will involve just one gate viz. the AO21 complex gate.

## 5. Simulation Results and Comparison

Semi-custom ASIC style implementation of homogeneous and heterogeneous 32-bit adders was considered in this work to make a performance comparison. The adders were physically realized using a high $V_t$ 32/28nm CMOS process [20]. The power, delay, and area estimates corresponding to the adders are given in Table 1. The design metrics estimated correspond to a typical case PVT specification with a recommended supply voltage of 1.05V and an operating junction temperature of 25ºC. For estimating the average power dissipation, about 1000 random input vectors were identically supplied to all the adders at time intervals of 5ns (200MHz) through a common test bench. The value change dump files generated through the functional simulations were subsequently utilized to estimate the average power dissipation. The critical path delays and area occupancies were also estimated. Synopsys tools were used to estimate the design metrics. Since minimum-size gates of the digital standard cell library [20] were used for realizing the various adders, this paves the way for a straightforward comparison of their design metrics post-physical synthesis. The least values of the design metrics estimated are highlighted using the blackened boxes in Table 1. Also, in Table 1, the legends namely Adder1 to Adder12 have been used for the ease of referencing. We will refer to these legends to briefly discuss the results obtained.

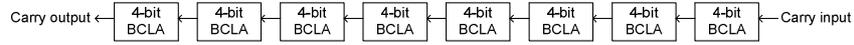

**Figure 4a. Homogeneous BCLA**

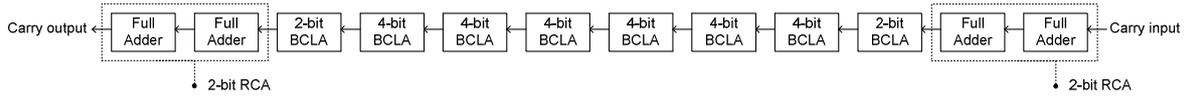

**Figure 4b. Hybrid BCLA-RCA**

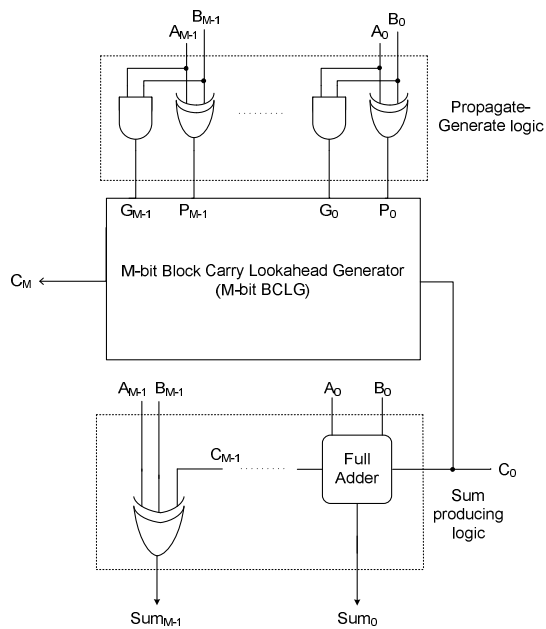

**Figure 4c. Block schematic of M-bit BCLA**

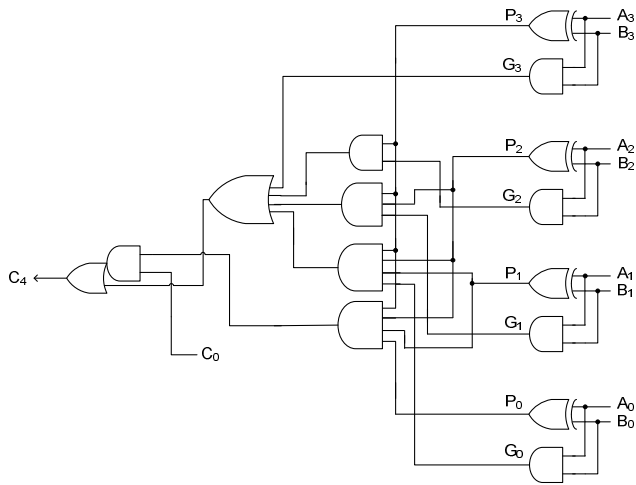

**Figure 4d. Gate-level detail of delay-optimized example 4-bit BCLG**

Besides the individual design metrics such as power, delay, and area, a combination of the design metrics such as the power-delay product (PDP), the energy-delay product (EDP), the area-delay product (ADP) and the power-delay-area product (PDAP) have also been used in the literature to serve as qualitative or nominal figure of merits to analyse the quality of results. Since the individual design metrics are desirable to be less, the combination design metrics are also desired to be less, which implies the least value of a combination design metric may point to the best optimized design. The combination design metrics were normalized before plotting their values for the various adders to maintain a uniformity. To perform normalization, the highest value of a combination design metric is considered, and it is used to divide the corresponding combination design metrics of all the adders. This implies the value of 1 represents the highest value of a combination design metric, which is not desirable, and the values lesser than 1 are desirable. The normalized PDP, EDP, ADP and PDAP plots corresponding to Adder1 to Adder12 (mentioned in Table 1) are portrayed in Figure 5.

**Table 1.** Design metrics of various 32-bit synchronous adders, estimated using a 32/28nm CMOS process

| Type of Adder | Adder Legend | Delay (ns) | Area ($\mu m^2$) | Power ($\mu W$) |
|---|---|---|---|---|
| RCA (Using the FA cell present in [20]) | Adder1 | 3.35 | 154.52 | 28.58 |
| RCA (XAC_FAs based on [5]) | Adder2 | 2.57 | 422.90 | 31.94 |
| RCA (XIMC_FAs based on [6]) | Adder3 | 2.74 | 365.97 | 25.06 |
| RCA (XOAC_FAs based on [7]) | Adder4 | 2.54 | 439.16 | 38.10 |
| RCA (DBFAs directly synthesized by Synopsys DC using [20]) | Adder5 | 2.38 | 357.83 | 26.66 |
| RCA (DBFAs based on [8]) | Adder6 | 1.58 | 569.28 | 42.34 |
| RCLA (Figure 3a) | Adder7 | 1.13 | 646.54 | 40.70 |
| Hybrid RCLA-RCA (Figure 3b) | Adder8 | 1.05 | 607.91 | 39.82 |
| BCLA (Figure 4a) | Adder9 | 1.26 | 500.16 | 43.80 |
| Hybrid BCLA-RCA (Figure 4b) | Adder10 | 1.12 | 457.97 | 41.72 |
| CSLA (Input partition of 8-7-6-4-3-2-2, based on [2]) | Adder11 | 1.13 | 418.32 | 61.51 |
| CSLA-BEC (Same input partition including BEC converters based on [1]) | Adder12 | 1.28 | 459.49 | 52.12 |

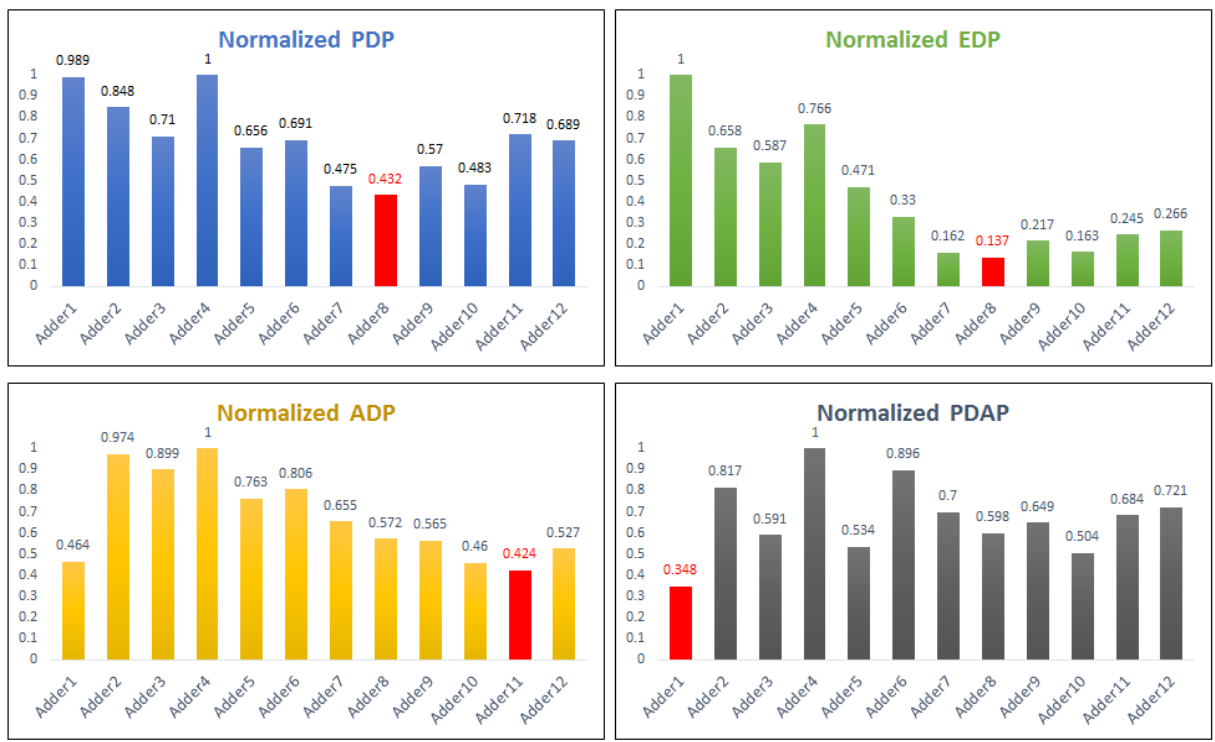

Figure 5. Normalized combination design metrics corresponding to various adders

In Figure 5, the least values of the corresponding combinational design metrics are highlighted using the red bar. It may be clear from Figure 5 that Adder8 (i.e. the hybrid RCLA-RCA) has the least PDP and EDP values, which are representative of low power design. ADP and PDAP are just nominal figure-of-merits. Based on the normalized ADP, Adder11 (i.e. the non-uniform input partitioned conventional CSLA) is the best among the adders considered. Although Adder1 (i.e. the RCA constructed using the full adder available in the standard digital cell library [20]) occupies the least area among all the adders, as noted from Table 1, the critical path delay of the RCA exceeds the critical path delay of the CSLA by 196.5%, and so the normalized ADP of Adder1 is greater than that of Adder11. Based on the normalized PDAP, Adder1 is found to be the best among the adders considered. With respect to area, the RCA architecture is the best, as observed in [21], and it is found to be the case here as well. However, from the perspective of low power, a hybrid RCLA-RCA architecture may be preferable.

As a future work, a comparison of the synchronous adders discussed here with the parallel-prefix synchronous adders [22] can be done. Moreover, the 64-bit addition operation could be considered as a natural extension of this work. Further, multi-operand addition operations, which are common in digital signal processing, can be considered and the performance of the various synchronous adders may be benchmarked based on those operations.